\documentclass[onecolumn,usenatbib,amsmath,amssymb,amsbsy]{mn2e}

\usepackage{natbib}
\usepackage{graphicx,amsmath,amssymb}

\def\eg{{\it e.g.}} 
\def\etal{{\it et al.}} 
\def\ie{{\it i.e.}}

\newcommand{\ncd}{\newcommand}
\ncd{\beq}{\begin{equation}}
\ncd{\eeq}{\end{equation}}
\ncd{\lbeq}[1]  {\label{eq: #1}}
\ncd{\refeq}[1] {(\ref{eq: #1})}
\ncd{\mrm}    {\mathrm}
\ncd{\der}{{\mathrm{d}}}
\ncd{\rtil}{\tilde{r}}
\ncd{\Mr}{\frac{2M}{r}}
\ncd{\rhotil}{\tilde{\rho}}
\ncd{\rstar}{r_{*}}
\ncd{\dell}{\partial}
\ncd{\mnote}[1]{\marginpar{\small #1}}

\ncd{\ec}{\check\epsilon}
\ncd{\rhoc}{\check\rho}
\ncd{\muc}{\check\mu}
\ncd{\pc}{\check{p}}
\ncd{\Gc}{\check\Gamma}
\ncd{\Oc}{\check\Omega}
\ncd{\betac}{\check\beta}

\ncd{\bldeta}{\boldsymbol{\eta}}
\ncd{\bldone}{\mathbf{1}}
\ncd{\blds}{\mathbf{s}}
\ncd{\bldk}{\mathbf{k}}
\ncd{\blde}{\mathbf{e}}

\ncd{\abs}[1] {|#1|} 
\ncd{\ubold}{\mathbf u}
\ncd{\Abold}{\mathbf A}
\ncd{\Bbold}{\mathbf B}
\ncd{\Mbold}{\mathbf M}
\ncd{\tsfrac}[2]{{\ts\frac{#1}{#2}}}
\ncd{\lagom}{\hspace{.6pt}}
\ncd{\muk}{k}
\ncd{\lagomdot}{{\mbox{\large$\cdot$}}}

\ncd{\stil}{\tilde{s}}
\ncd{\ftil}{\tilde{f}}
\ncd{\Otil}{\tilde{\Omega}}

\ncd{\ela}{\left(1-\frac{2m}{r}\right)}
\ncd{\nfrac}[2]{\left(\frac{n_{#1}}{n_{#2}}\right)^2}

\ncd{\shm}{S}
\ncd{\shmtwoD}{\mathcal{\shm}}

\ncd{\aaa}    { Astron.\ Astrophys.}
\ncd{\aip}    { Adv.\ Phys.}
\ncd{\adm}    { adv.\ math.}
\ncd{\am}     { Ann.\ Math.}
\ncd{\apl}    { Ann.\ Phys.\ (Leipzig)}
\ncd{\apny}   { Ann.\ Phys.\ (N.Y.)}
\ncd{\apj}    { Astrophys.\ J.}
\ncd{\apjl}    { Astrophys.\ J.\ Lett.}
\ncd{\cjp}    { Can.\ J.\ Phys.}
\ncd{\cmp}    { Commun.\ Math.\ Phys.}
\ncd{\cqg}    { Class.\ Quantum Grav.}
\ncd{\grg}    { Gen.\ Rel.\ Grav.}
\ncd{\ijmpd}  { Int.\ J.\ Mod.\ Phys.\ D}
\ncd{\ijtp}   { Int.\ J.\ Theor.\ Phys.}
\ncd{\invm}   { Invent.\ Math.}
\ncd{\jm}     { J.\ Math.}
\ncd{\jmaa}   { J.\ Math.\ Anal.\ Appl.}
\ncd{\jmp}    { J.\ Math.\ Phys.}
\ncd{\jpa}    { J.\ Phys.\ A}
\ncd{\lr}    { Liv.\ Rev.\ Rel.}
\ncd{\mnras}  { Mon.\ Not.\ R.\ Ast.\ Soc.}
\ncd{\mpl}   { Mod.\ Phys.\ Lett.} 
\ncd{\mpla}   { Mod.\ Phys.\ Lett.\ A} 
\ncd{\nature} { Nature}
\ncd{\nc}     { Nuovo Cim.}
\ncd{\npb}    { Nuc.\ Phys.\ B}
\ncd{\ph}     { Physica}
\ncd{\pla}    { Phys.\ Lett. A}
\ncd{\plb}    { Phys.\ Lett. B}
\ncd{\pr}     { Phys.\ Rev.}
\ncd{\pra}    { Phys.\ Rev.\ A}
\ncd{\prb}    { Phys.\ Rev.\ B}
\ncd{\prc}    { Phys.\ Rev.\ C}
\ncd{\prd}    { Phys.\ Rev.\ D}
\ncd{\prep}   { Phys.\ Rep.}
\ncd{\prl}    { Phys.\ Rev.\ Lett.}
\ncd{\prsla}  { Proc.\ Roy.\ Soc.\ Lond.\ A}
\ncd{\ptp}    { Prog.\ Theor.\ Phys.}
\ncd{\ptps}   { Prog.\ Theor.\ Phys.\ Suppl.}
\ncd{\rmp}    { Rev.\ Mod.\ Phys.}
\ncd{\spj}    { Sov.\ Phys.\ JETP}

\ncd{\ca}{\mathcal{C}_1}
\ncd{\cb}{\mathcal{C}_2}
\ncd{\w}{W}
\ncd{\betas}{\beta_{\star}}
 
\begin{document}

\title[Neutron Star Asteroseismology]
{ Neutron Star Asteroseismology. Axial Crust Oscillations in the 
Cowling Approximation.}

\author[Samuelsson \& Andersson]
{Lars Samuelsson\thanks{E-mail: lars@soton.ac.uk} and Nils Andersson\thanks{E-mail: na@maths.soton.ac.uk} \\
School of Mathematics, University of
Southampton, Southampton SO17 1BJ, UK}

\maketitle

\begin{abstract}
Recent observations of quasi-periodic oscillations in the aftermath of
giant flares in soft gamma-ray repeaters suggest a close coupling
between the seismic motion of the crust after a major quake and the
modes of oscillations in a magnetar. In this paper we consider the
purely elastic modes of oscillation in the crust of a neutron star in
the relativistic Cowling approximation (disregarding any magnetic
field). We determine the axial crust modes for a large set of stellar
models, using a state-of-the-art crust equation of state and a wide
range of core masses and radii. We also devise useful approximate
formulae for the mode-frequencies. We show that the relative crust
thickness is well described by a function of the compactness of the star
and a parameter describing the compressibility of the crust
only. Considering the observational data for SGR 1900+14 and SGR
1806$-$20, we demonstrate how our results can be used to constrain the
mass and radius of an oscillating neutron star.
\end{abstract}


\section{Introduction}
\label{sec:intro}
Asteroseismology aims to probe stellar physics via various observed
modes of vibration. To be successful in this endeavour one needs both
accurate observations and detailed theoretical models to test the
observations against. In the case of neutron stars we have until very
recently had neither reliable observations nor detailed theoretical
models. The situation appears to have changed with the observations of
quasiperiodic oscillations (QPOs) following giant flares in three soft
gamma-ray repeaters (SGRs)
\citep{israel:qpo,sw:qpo,sw:flare3,sw:flare2}. Analysis of the
relevant X-ray data has unveiled a number of periodicities, with
frequencies that agree reasonably well with the expected torsional
modes of the neutron star crust [\eg\ \citep{duncan:1998A}]. These
observations are tremendously exciting because they will allow us, for
the very first time, to test our neutron star oscillation models. They
also provide strong motivation to improve these models. After all, in
order to be accurate enough to make a comparison to the observations
meaningful, a model must be fully relativistic. It should also allow
for the presence of a strong magnetic field and possible superfluid
components.

In this paper we take some modest steps towards more realistic models
of pulsating neutron stars by including the crust elasticity in a
relativistic calculation of axial oscillations. To simplify the
problem we work within the Cowling approximation, \ie\ we neglect the
dynamical nature of spacetime. This is expected to be an accurate
approximation for axial modes in the crust since they originate in a
region of relatively low density and should not lead to significant
variations in the gravitational field. We are not considering the
class of axial gravitational-wave $w$-modes at this point
\citep{kokkotas:axialmodes}, apart from formally noting how they would
appear in our formulation (see Appendix~A). We calculate axial modes
for an up-to-date crust equation of state (not very controversial) for
a large set of core masses and radii, corresponding to different
(unspecified) supranuclear equations of state in the core and
(similarly unspecified) central pressures. Our approach allows us to
largely ignore the detailed structure of the core which is only poorly
constrained by observations and theoretical considerations. We show
how, when used in conjunction with observed mode frequencies, our
numerical results can be used to put constraints on the global
equation of state. We also determine an approximate solution to the
mode-problem. This leads to useful formulae that show how the axial
modes depend on the key parameters, the star's mass and radius, the
crust compressibility and the shear modulus. When compared to our
numerical data these approximations are surprisingly good. We show how
they can be used to deduce the main stellar parameters from a set of
observed frequencies, thus outlining a complete asteroseismology
analysis.

Our study does not at this point account for the presence of a
magnetic field. This obviously means that one should be careful before
assuming that our results can be used to interpret the magnetar
observations. However, we feel that the magnetic problem is still
somewhat beyond reach. The main reason for this is the strong coupling
between any motion in the crust and Alfv\'en waves in the core. As we
have recently argued \citep{gsa:mhd} [see also
\citet{levin:magnetars}], this coupling is likely to lead to global
oscillations which involve significant core motion. This problem is
computationally orders of magnitude more difficult because the
magnetic field plays an active role. Having said that, it may well
turn out to be the case that the modes that dominate the observed
signal remain quite close to the pure crust modes, despite the
presence of the magnetic field. One can argue that this should be the
case if the modes are excited by crust ``cracking''
\citep{duncan:1998A}, where the bulk of the energy is deposited into
elastic motion \citep{gsa:mhd}. This is, of course, a qualitative
argument and better calculations are needed to establish whether or
not it is quantitatively useful.

Before we proceed, it is worth noting that the approach to neutron
star asteroseismology described in this paper is directly applicable
to weaker magnetic field pulsars provided that they are not too
rapidly rotating. Glitches in these objects may excite torsional
oscillations in the crust, even though such oscillations are yet to be
detected.

\section{Perturbation equations}

Our study is motivated by a key question for both theorists and
observers: To what extent can we use the observed QPOs to probe neutron star
physics? We attempt to answer (at least partially) this question by
calculating axial oscillation modes for a realistic neutron star crust
using a fully relativistic theory of elasticity that was developed in
a recent series of papers
\citep{ks:relasticityI,ksz:stability,ks:exact,ks:relastaxial} based on 
the work of \cite{cq:elastica}. Since we want to emphasise the results
and the possible implications for asteroseismology rather than various
computational technicalities we will consider a simple neutron star
model that (we believe) captures most of the key features that govern
torsional crust oscillations. 

In our model the core has radius $R_c$ and mass $M_c$ but we keep
the equation of state (EOS) unspecified. We can do this since, within our approximations, 
there
is no coupling between the crust and the core fluid. The EOS of the crust is
described in detail by \citet{samuelsson:thesis} and its essential
parts are based on the work of \cite{hp:expmatter} and
\cite{dh:eos}. The shear modulus is taken from \cite{oi:shearmod}
and we assume that the temperature is zero.

We let the background star be static and
spherically symmetric. This means that the spacetime metric 
is given by
\beq
  \der s^2 = -e^{2\nu}\der t^2 + e^{2\lambda}\der r^2 + r^2(\der \theta^2 + \sin^2\theta\der\phi^2).
\eeq
The perturbations are treated in the Cowling approximation, \ie\ we
neglect metric perturbations. This should be a good approximation
provided that the oscillation modes are confined to the relatively low
density region represented by the crust.  The perturbation equations
for the crust, assuming perfectly elastic matter, are derived in
Appendix
\ref{sec:A}. These equations allow for the solid to be
anisotropic. The anisotropy manifests itself by the existence of two
different speeds of shear waves, denoted by $v_{r}$ (radially
propagating with polarisation in an angular direction) and $v_{t}$
(propagation in an angular direction, polarised in the mutually
orthogonal angular direction). Throughout the paper we will use a
subscript $r$ to mean radial and $t$ to mean tangential (to the
spherical symmetry surfaces). Similarly, there will be two different
pressures, $p_r$ and $p_t$. In the isotropic limit we denote the
quantities without subscripts, \ie\ $v=v_r=v_t$ and $p=p_r=p_t$.

As a further simplification, we assume that the core is unable to
support traction. This effectively isolates the crust, since we
neglect, \eg\ viscosity and magnetic fields. While the former should
be a reasonable approximation, the latter is not valid in the context
of magnetars. As we have argued elsewhere \citep{gsa:mhd},
the crust-core coupling time scale in a magnetar is comparable to the
oscillation period of the crustal modes which implies that one ought
to consider global MHD modes. As this problem is seriously
challenging, we prefer to focus on the non-magnetised case here. After
all, if we take the toy-model discussed by \cite{gsa:mhd} at face
value then it seems plausible that the modes that are mainly excited
by crust cracking in the magnetic problem may be close to the pure
crust modes determined in the non-magnetic case.


In Appendix \ref{sec:A} we show that the axial perturbation equations
for an elastic solid in the Cowling approximation can be written
\begin{align}\lbeq{perteq}
  F'' + A'F' + BF = 0,
\end{align}
where a prime denotes differentiation with respect to the
Schwarzschild radius $r$ and
\begin{align}
  e^A &= r^4e^{\nu-\lambda}(\rho+p_t)v_{r}^2, \\
  B &= \frac{e^{2\lambda}}{v_{r}^2}\left[e^{-2\nu}\omega^2-\frac{v_{t}^2(l-1)(l+2)}{r^2}\right].
\end{align}
Here $F$ describes the amplitude of the fluid oscillations, $\rho$ is
the energy density, the integer $l$ is the usual angular separation
constant that enters when the perturbation variables are expanded in
spherical harmonics $Y_l^m(\theta, \phi)$ and $\omega$ is the
angular frequency.

In the case that we are considering, 
the boundary conditions require that the traction
vanishes at the top and the bottom of the crust. In terms of our
independent variable $F$ this leads to
\beq
  T=e^A F'=0 \quad \mbox{at $r=R_c$ and $r=R$}.
\eeq
It is convenient to use $T$ to reduce \refeq{perteq} to a first order system,
\begin{align}\lbeq{peqsys}
  T' &= -e^ABF, \\
  F' &= e^{-A}T. \lbeq{peqsys2}
\end{align}
This system is suitable for numerical integration since there are no
derivatives of the equation of state parameters, \eg\ the shear
modulus or the density (which are only known in tabular form). In the
isotropic limit these perturbation equations are equivalent to \eg\ those used by
\cite{yl:nonrad}.

We have used these equations to calculate axial crust modes for a
large number of stellar models with different core masses and radii, implicitly
corresponding to a variety of supranuclear equations of
state in the neutron star core. We want to use these results to
develop a workable strategy for asteroseismology. That is, we would
like to i) identify the key parameters that govern the various modes,
and ii) try to represent the results in such a way that a parameter
``inversion'' becomes possible given actual observations. The basic
idea behind this kind of analysis has already been discussed for
neutron star oscillations that radiate gravitational waves at a
significant level \citep{ak:seismo}. Since the analysis is much aided
by ``analytic'' formulae, we will develop a suitably simple
approximation scheme for the axial modes. This will lead to a set of
empirical expressions that can be used to address the inverse
problem.

\section{Simple analytic approximations}

We want to find an approximation to the axial crust modes. 
To do this, we consider Eq.\ \refeq{perteq} and introduce a new
Regge-Wheeler type radial coordinate $x$ through
\beq
  \frac{dx}{dr} = e^{-A}.
\eeq
The perturbation equation then becomes 
\beq
  \frac{d^2F}{dx^2}+e^{2A}BF=0,
\eeq
i.e., it is written on a form that lends itself
to a WKB-type approximation. Since we are interested in 
making progress analytically we consider only the
lowest order approximation. This means that we assume that the 
solution can be written
\beq
  F=\ca e^{i\w(x)} + \cb e^{-i\w(x)}, \qquad \w(x)=\int_{R_c}^x e^{A}B^{1/2} dx = \int_{R_c}^r B^{1/2} dr.
\eeq
At the base of the crust ($r=R_c$) 
we need to ensure the vanishing of the traction. Hence, we impose
the boundary condition
\beq
  F' = iB^{1/2}(\ca-\cb)=0 \quad \Rightarrow \quad \ca=\cb.
\eeq
Meanwhile, the analogous condition at the surface ($r=R$) implies that
\beq \lbeq{bcR}
  F'=i\ca B^{1/2}(e^{i\w(R)}-e^{-i\w(R)})=0 \quad \Rightarrow \quad \w(R)=\int_{R_c}^R B^{1/2} dr = n\pi.
\eeq

To make further progress, we next make the approximation that the
shear speeds are constant (an approximation which is good
throughout much of the crust) and that $\nu$ and $\lambda$ are
constant (which is a reasonable approximation since the crust mass in
negligible compared to that of the core).
Then, assuming that 
\beq \lbeq{expcond}
  \omega^2\gg e^{2\nu}v_t^2\frac{(l-1)(l+2)}{r^2},
\eeq
we may Taylor expand $B^{1/2}$ to get
\beq
  B^{1/2}\approx e^{\lambda-\nu}\frac{\omega}{v_r}\left[1-\frac{e^{2\nu}v_t^2(l-1)(l+2)}{2\omega^2r^2}\right],
\eeq
which may be integrated to yield, using Eq. \refeq{bcR},
\beq
  \omega^2 - e^{\nu-\lambda}\frac{n\pi v_r}{\Delta}\omega -e^{2\nu}\frac{v_t^2(l-1)(l+2)}{2RR_c} \approx 0,
\label{app_n}\eeq
where $\Delta=R-R_c$. This equation provides a useful first approximation to the frequencies of 
the axial crust modes. 

Before considering the overtones, let us discuss the case of the 
fundamental crust modes, which correspond to $n=0$. 
For this case we immediately find
\beq
  \omega^2 \approx \frac{e^{2\nu}v_t^2(l-1)(l+2)}{2RR_c} \qquad (n=0).
\label{fund}\eeq
However, this result shows that our approximation is not consistent
for the fundamental mode. It is clear that the condition \refeq{expcond} is 
not satisfied throughout the crust. Nevertheless, our approximate result provides useful insights
into the scaling with various parameters. We can compare it to, for
example, the formula used by \cite{piro:flares} [his Eq. (9)]. This is the
standard estimate, which is arrived at via a plane-wave approximation,
and it can be written
\beq
  \omega^2 \approx \frac{e^{2\nu} v^2l(l+1)}{R^2},
\eeq
where $v$ is the isotropic speed of shear waves.
This result differs from ours in two important ways. First of all, the 
 $l$ dependence is different. Secondly, our formula replaces 
$R^2$ by $RR_c$.  

\begin{figure}
\centerline{
{\includegraphics[width=0.45\columnwidth,angle=0,clip]{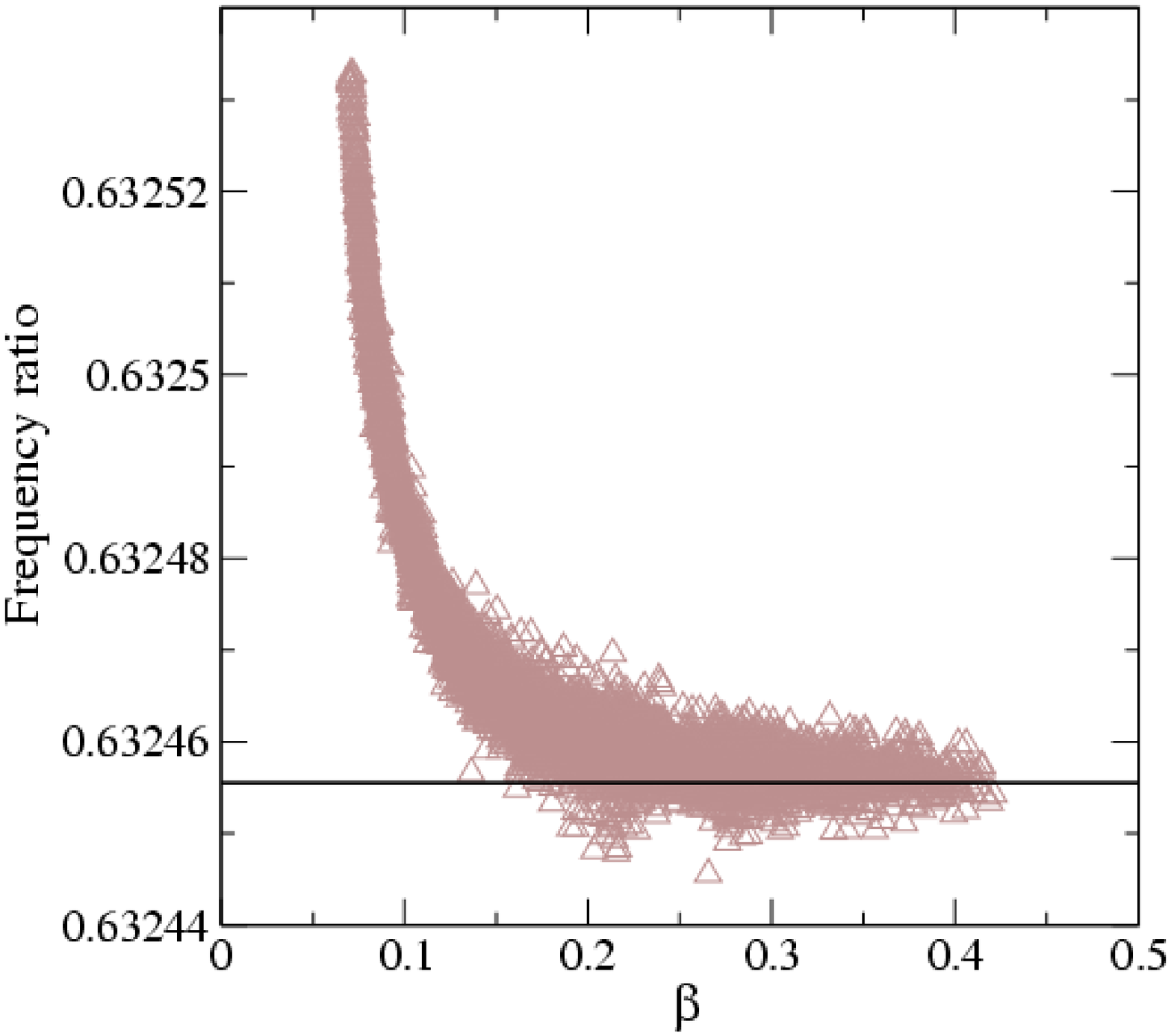}}
{\phantom{M}}
{\includegraphics[width=0.41\columnwidth,angle=0,clip]{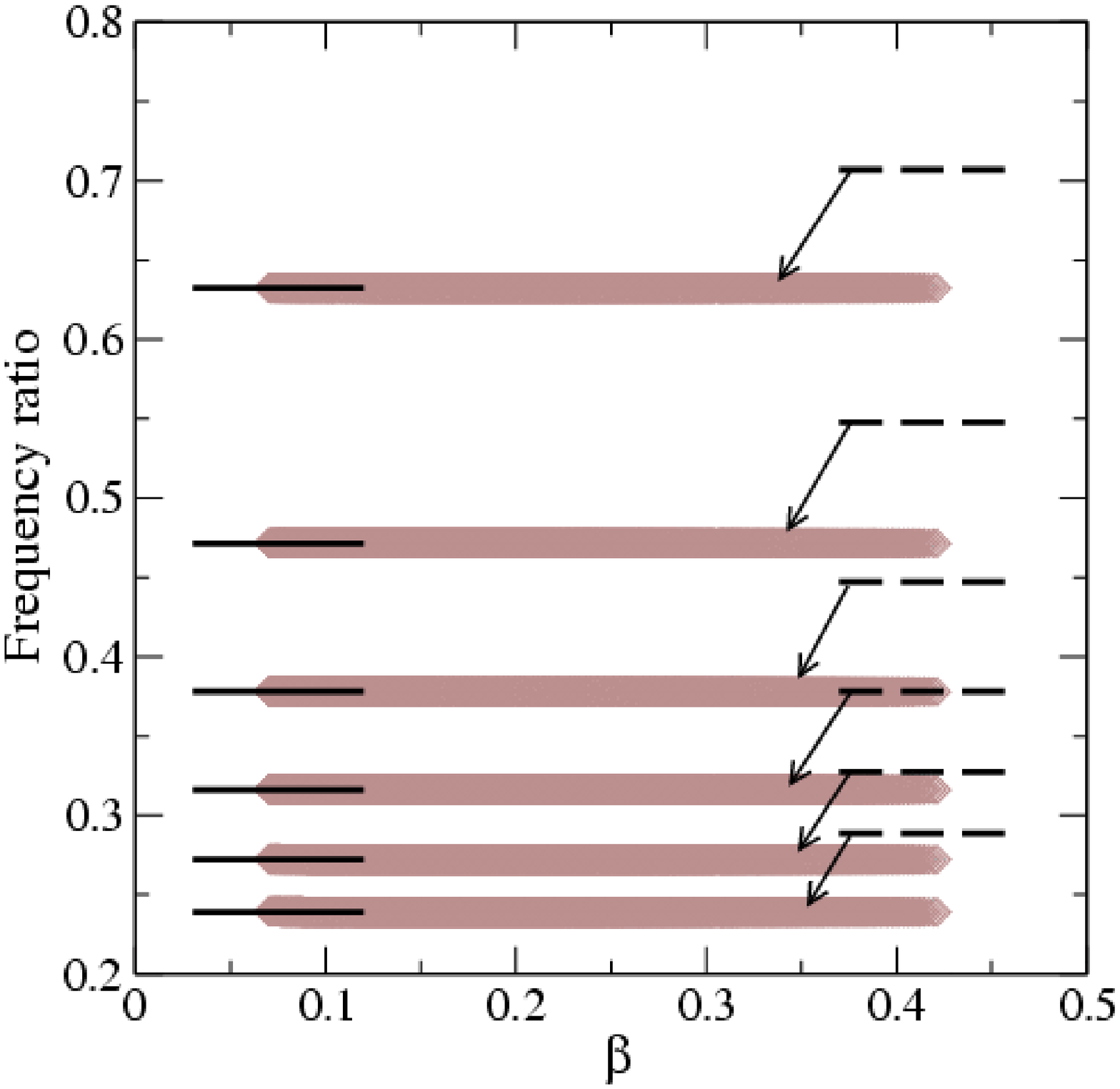}}
}
\caption{In the left panel we show that the ratio of the frequencies 
  of the fundamental modes for $l=2$ and $l=3$ agrees well with that
  predicted by our $(l-1)(l+2)$ scaling which in this case gives a
  ratio of $\sqrt{5/2}\approx 0.6324555$ (shown as a solid line). The
  small deviation seen for low $\beta = M/R$ is typical for the first
  few $l$'s. The remarkable agreement for typical neutron star values
  $\beta \approx 0.2$ should be noted. It is also worth remarking that
  our numerical code has an accuracy of a few times $10^{-5}$, which
  means that some of the remaining spread in this range may be due to
  numerical error. In the right panel we show that the numerically
  computed ratio of the frequencies of the fundamental modes agrees
  well with that predicted by our $(l-1)(l+2)$ scaling (shown on the
  left as solid lines). The $l(l+1)$ prediction is show on the right
  as dashed lines.  Of course, for large $l$ (so that the plane wave
  approximation is valid) we get $(l-1)(l+2)=l^2+l-2\rightarrow
  l(l+1)$. Nevertheless, the discrepancy for the first few values of
  $l$ may lead to an erroneous multipole assignation for a given
  observed mode-signal. }
\label{fig:frat}
\end{figure}

It is easy to demonstrate that the difference in the $l$-dependence is
important, and that our formula represents the correct behaviour. In
order to do this, let us consider the ratio between the fundamental
quadrupole ($l=2$) mode frequency and various higher multipoles. In
the ratio the dependence on, for example, the shear wave speed
disappears and we are left with a simple scaling with $l$. If we also
compare the approximate formula to the results obtained from solving
the full problem numerically we get insight into the accuracy of some
of our assumptions, in particular whether it is reasonable to treat
the shear wave speed as if it were constant and effectively ``the
same'' for the different modes. This comparison is illustrated in
figure~\ref{fig:frat}. The left panel shows the ratio between the
fundamental $l=2$ and $l=3$ modes for our approximate formula and our
total sample of  9000 numerical models with parameters in the
range $8\le R_c \le 16$ km and $0.8\le M_c/M_\odot \le 2.3$. From the
figure it is clear that the numerical results scale with $l$ in the
way suggested by our (inconsistent) approximation. This is further
demonstrated by the data in the right panel of figure~\ref{fig:frat},
which extends the comparison to higher multipoles. This figure 
shows that an assumed scaling with $l(l+1)$ is likely to lead to an
erroneous identification of the various higher multipoles that may be
present in an observed signal. It may, for instance, be noted in table
\ref{tab:f} that we arrive at a different multipole
identification compared to \cite{sw:qpo} for SGR 1900+14 and
\cite{israel:qpo} for SGR 1806$-$20.
  
Let us now consider the various overtones $n\neq0$ for any given $l$.
We then need to solve the quadratic equation (\ref{app_n}). Expanding
the resulting square root in the small parameter $\Delta/R$ and
ignoring the negative root we get
\beq
  \omega \approx e^{\nu-\lambda}\frac{n\pi v_r}{\Delta}\left(1 +
    e^{2\lambda}\frac{(l-1)(l+2)}{2\pi^2}\frac{v_t^2}{v_r^2}\frac{\Delta^2}{RR_c}\frac1{n^2}\right).
\label{over}
\eeq
The second term in the parenthesis is negligible
for moderate $l$.  It is worth noting that it is $v_{t}$ that affects the
$n=0$ modes whereas the $n>0$ modes are primarily determined by
$v_{r}$. It should also be emphasised that condition \refeq{expcond}
holds for the overtones, which means that our approximation is
consistent in this case.

In order to estimate the overtone frequencies for any given 
$M$ and $R$ (say) we need to provide the crust thickness $\Delta(R,M)$ which 
then leads to $R_c$. In Appendix \ref{sec:B} we
show that the crust thickness is well approximated by 
\beq\lbeq{Delta}
  \frac{\Delta}{R} \approx \left(\frac{\beta}{\alpha}e^{2\lambda}+1\right)^{-1},
\eeq
where $\beta=M/R$ is the stellar compactness and $\alpha$ is a
parameter that depends on the equation of state and which essentially
measures some average compressibility of the crust. As discussed in
Appendix~B, the relevant value for our crust equation of state is $\alpha=0.02326$.

We want to combine these various approximations to get explicit
expressions for the mode frequencies that can be used to compare to
either numerical results for detailed relativistic models or
observational data. To do this, we first assume that the crust is
isotropic by setting $v_r=v_t=v$. If we then compare Eq.~(\ref{fund})
and (\ref{over}) to our full numerical results we find that they
provide reasonable approximations for \emph{the same value} of the
shear wave speed $v$, provided that we adjust the factor of $1/2$ in
(\ref{fund}). Since we already knew that Eq.~(\ref{fund}) was not
derived in a consistent way, this manipulation does not seem
unreasonable.

The shear speed varies very little in the crust and has a value $\sim
10^8$ cm/s. Writing $v = v_0\times 10^8$ cm/s, where $v_0$ is a
dimensionless parameter we may write our frequency estimates as
\begin{align}
 f &\approx 11.85\frac{v_0}{R_{10}}\frac{\sqrt{(l-1)(l+2)}}{2}
    \sqrt{\frac{(1.705-0.705\betas)(0.1055+0.8945\betas)}{\betas}}
    , \quad  (n=0) \\
 f &\approx 473.77 n \frac{v_0}{R_{10}}(0.1055+0.8945\betas), \quad (n>0) 
\end{align}
where $R_{10}=R/10$~km and $\betas=\beta/0.2068$ ($\beta_\star=1$
for a star with radius 10~km and mass $1.4 M_\odot$, obtained using
the most recent value for $G$ etc.).  Comparing to our numerical data
we find that $v_0\approx 2.34$. This value may appear to be
surprisingly large given that, for our EOS, the maximum shear speed is
about $2\times10^8$ cm/s. However, it should be remembered that the
parameter $v_0$ is obtained through comparison with the full numerical
solutions and therefore represent some ``weighted average'' shear
speed. In this (implicit) averaging geometrical factors and explicit
EOS dependence may well contribute factors of order unity. In a sense,
the replacement of $R$ with the geometrical mean radius of the crust
$\sqrt{RRc}$ represent the first order geometrical effect.  It is also
worth pointing out that the simple linear scaling with $n$ does not
represent the behaviour found in the numerical study. Again, this is
not surprising as the eigenfunctions will have different forms and
hence depend more strongly on the EOS in different parts of the
crust. In general the mode frequencies seem to grow less rapidly than
linear as a function of $n$.  Another interesting point to note is the
absence of an overall redshift factor in the estimated overtone
frequencies. This somewhat counter-intuitive result arises since the
redshift factor in (\ref{over}) is essentially cancelled by the
corresponding factor in the formula for the crust thickness.

\begin{figure}
  \centerline{
    \includegraphics[width=0.45\columnwidth,angle=0,clip]{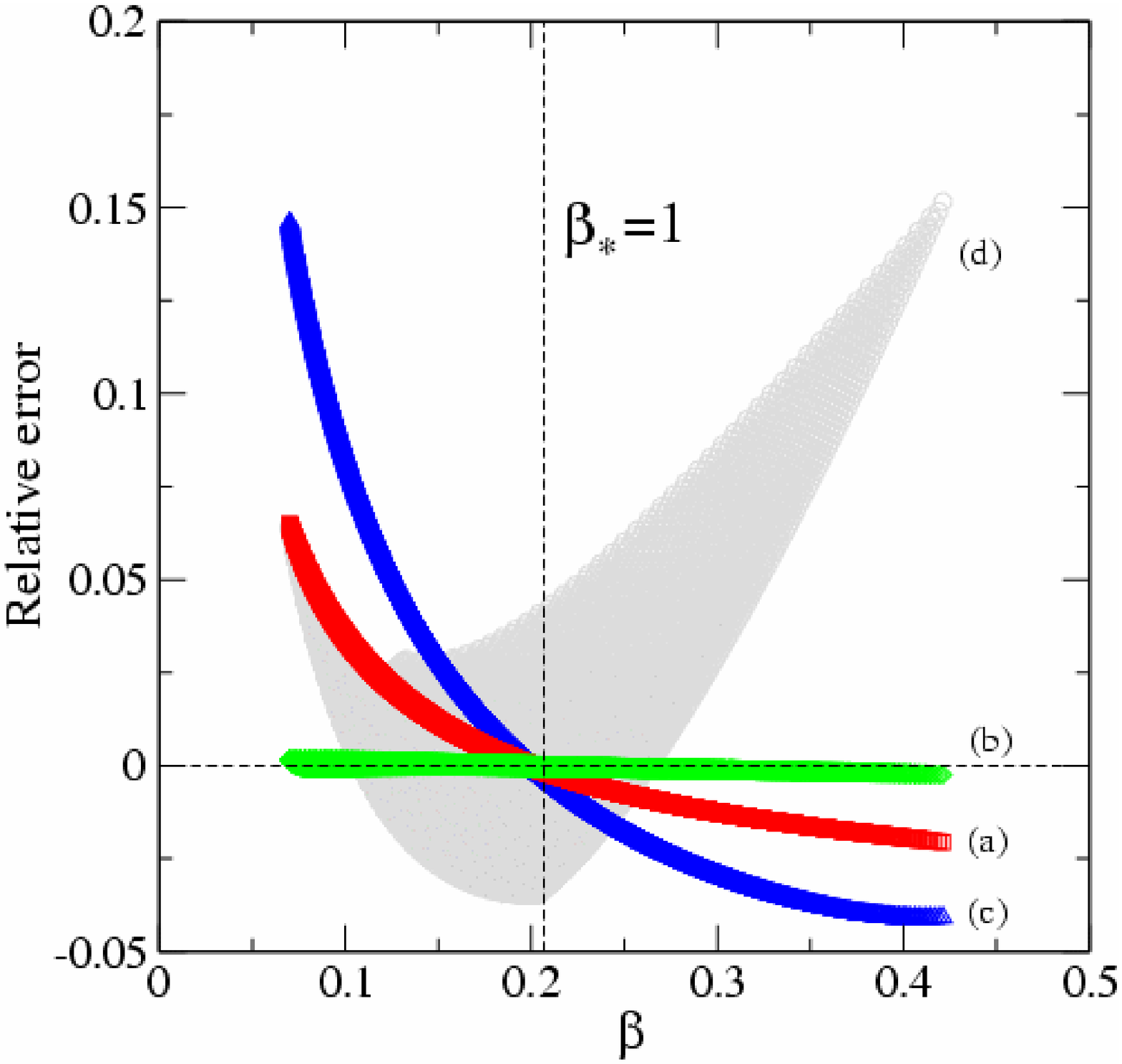}
    \phantom{M}
    \includegraphics[width=0.45\columnwidth,angle=0,clip]{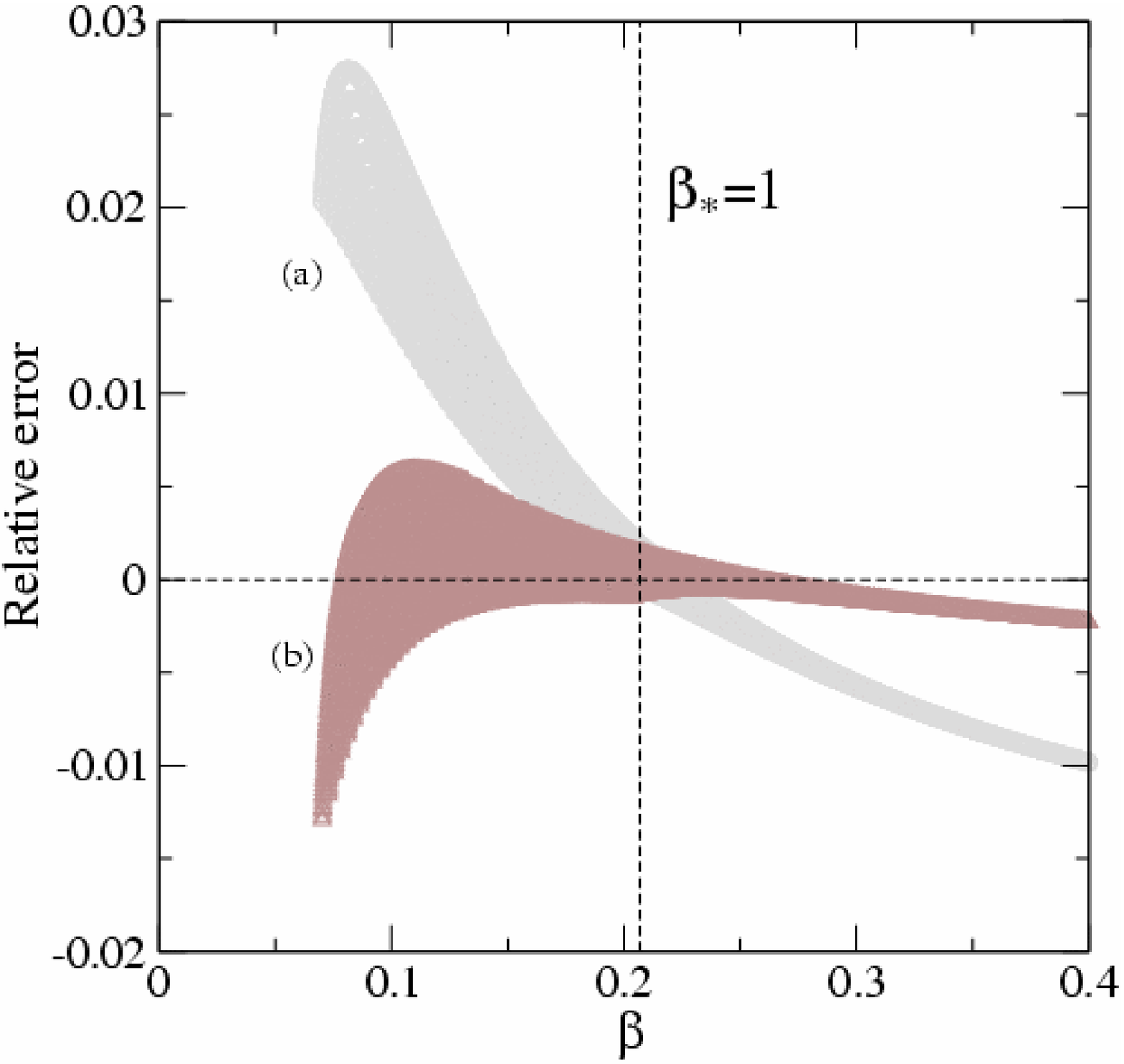}
  }
\caption{In the left panel we compare various estimates for the fundamental $l=2$ torsional mode. 
 We show the relative error, compared to numerical data, of the
 frequency estimates as functions of $\beta$. In each case we have
 rescaled the estimates to give approximately the numerically computed
 value of the frequency at $\betas=1$ (see the main text). This
 corresponds to choosing an appropriate value of the shear
 speed. Comparing Piro's result (c) with ours (a) we see that the
 $(RR_c)^{-1/2}$ scaling performs better than the standard
 $R^{-1}$. We also note that the extrapolated ``general relativity
 correction factor'' of Duncan (d) not only makes the estimate less
 accurate, but actually destroys the clean $\beta$-dependence. Also
 shown is our fit of the frequencies (b) which is accurate to within
 0.2 \%. In the right panel we show the relative error in the estimate
 for the frequency of the first overtone both with (b) and without (a)
 the empirical fitting factor.}
\label{fig:cmpf}
\end{figure}

The accuracy of our approximate formulae is illustrated in Figure~\ref{fig:cmpf}.
When comparing the approximations to the numerical data we
note that, for the fundamental modes, the error is a smooth function of
$\beta$. The situation therefore lends itself to improvement by
fitting. Playing the same game with the overtones we find
\begin{align}
 f &\approx 27.65\frac{1}{R_{10}}\frac{\sqrt{(l-1)(l+2)}}{2}\sqrt{\frac{(1.705-0.705\betas)(0.1055+0.8945\betas)}{\betas}}
  \frac{\betas}{1.0331\betas-0.0331} , \quad  (n=0)  \lbeq{f0est}\\
 f &\approx 1107.3 \frac{1}{R_{10}}n(0.1055+0.8945\betas)\frac{\betas}{1.0166\betas-0.0166}, \quad (n>0) \lbeq{f1est}
\end{align}
These frequency estimates are evaluated against numerical data in figure \ref{fig:cmpf}.
We see that they perform extremely well, with relative errors typically smaller  than 1\%.
Clearly, our various approximate formulae provide a very accurate representation of the 
full numerical results. We therefore anticipate that they
will prove useful for any attempts to 
infer physical parameters from observational data.

\section{Seismology}

In this section we discuss how one can use our numerical and
approximate results for axial modes in the neutron star crust to infer
physical parameters from observations, eg. the QPOs found in the tails
of magnetar flares. To carry out this analysis we assume that the
observed QPOs in the frequency range $\sim 28 - 160$ Hz are connected
with fundamental ($n=0$) elastic modes with different $l$'s, whereas
the QPO at $\sim 626$ Hz is an overtone (see below for a brief summary
of the observations). It is important to note that these assumptions
may not be valid, especially since one can argue that the crust motion
excites motion also in the core fluid via the magnetic field coupling
\citep{gsa:mhd,levin:magnetars}. If this is the case, then the mode
problem that we have solved in this paper does not provide a complete
picture. However, even though our analysis may not be entirely
realistic in this sense we believe that our discussion provides a
useful example of the strategy.  We should also stress that, until
fully relativistic modes for magnetised stars are calculated, this
analysis may be the best one can expect.

\subsection{Observations}

Before considering the inverse problem of finding global neutron star
properties from observed spectra let us briefly discuss the
observations. So far there have been three recorded giant flares, all
associated with SGRs.

\paragraph*{SGR 0526$-$66:}
This flare took place in March 5 1979 \citep{Mazets:flare1979}. Only a
weak QPO was detected at $\approx 44.5$ Hz \citep{barat:qpo1979}. Due
to the meagre data and the fact that this feature was detected in the
peak of flare with significant dead-time effects in the data we will
not consider this observation further.

\paragraph*{SGR 1806$-$20:}
Erupted December 27 2004 \citep{hurley:flare1806,palmer:flare1806}.
\cite{israel:qpo} detected QPOs at $18.1\pm 0.3$,
$30.4\pm 0.3$ and $92.5$ Hz in the data from the \emph{Rossi X-ray
Timing Explorer (RXTE)}. 
\cite{sw:flare2} later
examined the data from the \emph{Ramaty High Energy Solar
Spectroscopic Imager (RHESSI)} and found QPOs at $17.9\pm 0.1$,
$25.7\pm 0.1$, $30$ (weaker), $92.7\pm 0.1$ and $626.46\pm 0.02$
Hz. \cite{sw:flare3} have recently detected a feature at $\approx 150$
Hz\footnote{This feature was brought to our intention via private
communication with Watts. After this work was completed
\cite{sw:flare3} published their results from a reanalysis of \emph{RXTE} 
data. In addition to the new 150 Hz feature several higher frequency
QPOs were found. These have not been considered in this work.} as well
as showing that the $\approx$ 30 Hz QPO is better fitted by 29 Hz. In
our analysis we exclude the lower frequency QPOs. This choice is
guided by our recent toy problem
\citep{gsa:mhd}. Hence, we consider the 29, 92.7, 150.3 and 626 Hz
modes. The inclusion of the 150 Hz QPO does not change the conclusions
substantially (which, incidentally, support the identification of the
29 Hz QPO with the $l=2$ mode and that our scaling with $l$ is the
correct one). In table \ref{tab:f} we display these frequencies
together with the modes that we identify them with.

\paragraph*{SGR 1900+14:}
A giant flare from this magnetar was detected on August 27 1998
\citep{hurley:flare1900,feroci:flare1900}. \cite{sw:qpo} found
evidence for QPOs at $28\pm 0.5$, $53.5\pm 0.5$, $84$, $155.1\pm 0.2$
Hz in the data from \emph{RXTE}. In our analysis we use all of these
frequencies. The modes that we identify them with are listed in
table
\ref{tab:f}.

\begin{table}
\centerline{ 
\begin{tabular}{|c|c|c|c|}
\hline
\multicolumn{2}{|c|}{SGR 1806$-$20} & \multicolumn{2}{|c|}{SGR 1900+14} \\
\hline
f [Hz] & mode & f [Hz] & mode \\
\hline
$29$             & ${}_0t_2$ &    $28\pm0.5$     & ${}_0t_2$ \\
$92.7\pm 0.1$    & ${}_0t_6$ &    $53.5\pm0.5$   & ${}_0t_4$ \\
$150.3$            & ${}_0t_{10}$ & $84$           & ${}_0t_6$ \\
$626.46\pm 0.02$ & ${}_1t_l$      & $155.1\pm 0.2$ & ${}_0t_{11}$ \\
\hline
\end{tabular}}
\caption{Observed QPO frequencies and suggested corresponding elastic crust modes. 
The identified modes are denoted according to {${}_nt_{l}$}. Note that
the true $l$-dependence differs from the standard {$l(l+1)$} which
explains why our identifications are different from those of
Strohmayer \& Watts (2005) (SGR 1900+14) and Israel \etal\ (2005) (SGR
1806$-$20). It may also be noted that the rather large allowed error
(6 \%) in our analysis makes it possible to identify the 150 Hz QPO in
SGR 1806$-$20 with both an {$l=9$} and and {$l=10$} mode in different
regions of the $M$-$R$ parameter space. However, the presence of the
overtone at 626 Hz marginally rules out the {$l=9$} option.}
\label{tab:f}
\end{table}

\subsection{Seismology with the analytic expressions}

It is straightforward to use the analytic approximations that we
derived in the previous section to analyse the observed frequencies
and extract the key neutron star parameters. 

Let us first assume that we observe the fundamental quadrupole mode
together with the first overtone $(n=1)$.  In the data for SGR 1806$-$20
it seems reasonable to assume that the first is represented by the
29~Hz QPO, while the latter corresponds to the 626~Hz mode. From our
approximate formulae we immediately see that the ratio of these modes
provides an expression that only contains the compactness
$\betas$. 
In effect, the data provides a curve in the $M-R$ plane on which the
true stellar model should lie. Solving this constraint for the
compactness we find that $\beta\approx 0.12$, \ie\ $R\approx
8.1M$. From Eq.~\refeq{Delta} we also see that the relative crust
thickness is $\Delta/R\approx 0.17$. We can then insert this value for
$\beta$ in the expression for (say) the fundamental mode. This allows
us to solve for the radius, and we find that $R\approx 11.4$ km, which
means that $M\approx 1.41$ km $\approx 0.96 M_{\odot}$ and
$\Delta\approx 1.9$ km. These values do not seem unreasonable, even
though the inferred mass is quite low and the crust surprisingly
thick. Most likely there are systematic effects due to the magnetic
field. These are, of course, not accounted for in our analysis.
 
Next consider the fundamental modes for different values of $l$. First
note that the scaling with $l$ allows us to immediately work out the
ratio of the different mode frequencies. Once we assign the
$\approx$30~Hz QPO to the fundamental $l=2$ mode we can infer the
multipoles of the various higher frequency modes in the observed
data. The conclusions of this analysis are given in Table~\ref{tab:f}.
Once we have determined the various multipoles, our approximate
formula for the fundamental modes can be used to constrain the stellar
parameters. Let us take the quadrupole mode as an example. If the
frequency is known, then the approximate formula gives the stellar
radius as a function of the compactness $\beta$. Since $M=\beta R$ we
again have a constraint curve in the $M-R$ plane. In order to be
consistent with the results for the $n=1$ overtone, the two curves
must intersect. The point of intersection immediately provides us with
the mass and radius of the star. Available results for higher
$l$-modes provide further constraints that can be used to verify the
consistency/accuracy of the inferred stellar parameters. Of course,
since our model is an idealisation (a spherical star with an isotropic
crust etcetera) one would expect the use of real data to lead to a
spread of the various curves in the $M-R$ plane.  This uncertainty can
to some extent be used to assess the reliability of the parameter
extraction.

\subsection{Using numerical results to identify parameters}

In our numerical code we solve the background Einstein equations
together with the perturbation equations, specifying the core mass and
radius. In the absence of discontinuous tangential pressure at the
crust-core interface neither this boundary nor the top of the crust
present any difficulties. [If there is a jump discontinuity in the
tangential pressure we have one more parameter in the problem, see
\cite{ks:relasticityI} and \cite{ksz:stability} for  discussions.] It is
therefore straightforward to integrate the equations both inwards and
outwards to a matching point where we make sure that the Wronskian of
the system \refeq{peqsys} and \refeq{peqsys2} vanishes. We have
checked that the results do not depend on the choice of matching
pressure.  The simple nature of our model allows us to quickly compute
many modes ($2\le l \le 14$, $0\le n \le 3$) for a large set of core
masses (45) and radii (200) numerically.

\begin{figure}
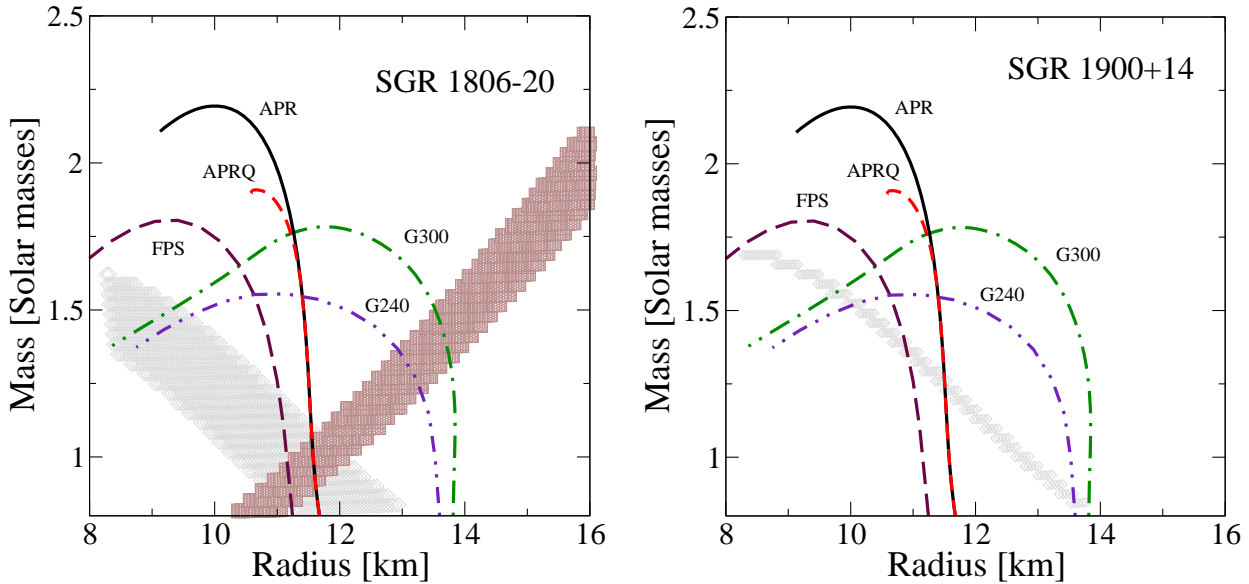

\centerline{\includegraphics[width=0.45\columnwidth,angle=0,clip]{xg1806.eps}
\phantom{M}
\includegraphics[width=0.45\columnwidth,angle=0,clip]{xg1900.eps}
}
\caption{Seismology analysis based on our numerical axial-mode results. 
For the flare in SGR 1806$-$20 (left panel) we associate the lower
frequency QPOs (approximately 29, 93 \& 150 Hz) with fundamental
($n=0$) modes with $l=2,6,9$ or 10 respectively. The inclusion of the
150 Hz mode does not change the picture substantially. The higher
frequency QPO (626 Hz) is associated with a $n=1$ mode of arbitrary
$l$. The models allowed by the lower QPOs form a rather broad region
that is orthogonal to the line corresponding to models that have an
$n=1$ overtone of the right magnitude.  As discussed in the main text,
the true stellar parameters should belong to the region of overlap.
For the flare in SGR 1900+14 (right panel) we use all the reported
QPOs (approximately 28, 53, 84 \& 155 Hz) and find that the sequence
of $l$'s is (2,4,6,11). The larger number of modes leads to a narrower
allowed region for the given tolerance. However, the lack of overtones
in the data means that the stellar parameters are much less
constrained in this case.  }
\label{fig:seismo}\end{figure}

Our aim is to use this data set to determine models that match all the
observed frequencies listed in table \ref{tab:f} (to some accuracy). As
indicated above, one has to allow for some level of uncertainty when
comparing the observed data to our numerical calculations. In our
analysis we have, rather arbitrarily, chosen to allow a relative
error of 6 \% in the frequencies.  For each of the two flares we
associate a subset of the observed QPO frequencies with axial
oscillation modes and search our computed data set for stellar models
which has all the required frequencies, for some sequence of
$l$'s. For SGR 1900+14 we use all four observed QPOs and infer that
the excited modes correspond to $l=2,4,6,11$. For SGR 1806$-$20 we
assume that the frequencies lower than about 30 Hz are not directly
associated with axial crust oscillations [see \cite{gsa:mhd} for a
possible explanation of these modes]. We are then left with a sequence
of three fundamental modes which we identify with the multipoles
$l=2,6,10$. The last one, when only compared to the other
fundamental modes, could also be allowed to have $l=9$ in parts of the
parameter space. However, for SGR 1806$-$20 we also a higher frequency QPO
which we associate with an $n=1$ overtone with arbitrary (but low)
$l$. This marginally restricts the higher fundamental mode to  $l=10$.

The results of the analysis are displayed in Figure~\ref{fig:seismo}.
In each panel we plot the mass-radius relation (as lines) for a few
proposed EOSs. The equations of state are chosen for illustration only
and include the A18-$\delta v$+UIX$^*$ model of
\cite{apr:eos} with and without a deconfined quark core (APR \& APRQ),
two examples from \cite{glendenning:compactstars} (G240 \& G300) and
the FPS EOS of \cite{pr:fps} (FPS). On top of this we display, as grey
diamonds, the allowed masses and radii if the given set of observed
low frequency QPOs are associated with elastic oscillations. In the
plot of SGR 1806$-$20 the models which have an $n=1$ overtone is shown
as dark squares.

The numerical results nicely illustrate the strategy for neutron star
seismology discussed for the approximate formulae. In particular, the
case of SGR 1806$-$20 indicates that the detection of overtone modes is
extremely valuable. Given both the fundamental mode and an overtone
the stellar parameters can be constrained to a relatively narrow
region of the $M$-$R$ plane. This, of course, then provides useful
constraints on the supranuclear core EOS.

\section{Discussion}

In this paper, we have considered torsional oscillation modes in
neutron star crusts. Using a state-of-art EOS in the crust but
otherwise simplifying the problem as far as possible we have shown how
global properties of the star may be deduced from observations of such
modes. We have done this using both approximate formulae and a large
sample of fully numerical mode-results.  As an example we applied our
methods to SGRs assuming that the post-flare QPOs are connected with
pure crustal oscillations. The results show that the devised strategy
for asteroseismology is, in principle, quite powerful. Of course, the
basic model needs to be improved before the inferred stellar
parameters should be taken too seriously. In particular, we expect
systematic effects due to our neglect of the magnetic field, which
obviously should be important for magnetars. One can immediately think
of several ways to improve upon our calculations. Below we list the
most important.

For a magnetic star, one would not expect the QPOs to be \emph
{exactly} at the pure crustal frequencies. The toy model of
\cite{gsa:mhd} suggests that the QPOs are really global oscillations
of the entire star, but that the modes that are predominately excited
lie close to the pure crust modes.  In addition one would expect the
geometry and strength of the magnetic field to have a key role in
determining which modes are excited and how strongly. Clearly, this
question can only be settled in a realistic spheroidal model. Work on
that problem should be strongly encouraged.

The magnetic field will not simply lead to an offset of the mode
frequency with respect to that of a pure crust mode. The nature of the
modes will also be affected by the presence of the field. The
considerations of \cite{piro:flares} show that (in his plane-parallel
geometry) the fundamental modes are only weakly affected by the
magnetic field, whereas the overtones are strongly dependent on the
field. This is to be expected from our expressions \refeq{f0est} and
\refeq{f1est} where we see that the fundamental modes are determined
by the tangential shear velocity $v_t$ while the overtones instead
depend on the radial velocity $v_r$. A magnetic field will boost the
elasto-MHD speeds along the field direction and it therefore seems
possible that the toroidal part of the field will affect the
fundamental modes whereas the poloidal part will influence the
overtones. Should this prove to be true for the spherical case it
would be very interesting since it would offer a possibility of
extracting information concerning the field geometry. That the
geometry of the magnetic field configuration is essential to the
behaviour of the torsional modes was emphasised by \cite{mps:qpo}
and recently by \cite{sks:torsional}. 

In our model we have assumed zero temperature. 
A non-zero temperature will lead to a systematic effect since the shear
modulus will decrease \citep{oi:shearmod}. It will also mean that the
crust will be slightly thinner as the ocean is deeper. In our
calculations we have integrated the equations all the way to the
density of iron at zero pressure.

In the deep crust matter is not expected to be 
microscopically isotropic, but instead form so-called pasta phases. If
these anisotropies extend to macroscopic (fluid-dynamic) scales,
\ie\ if they do not average out over large distances compared to the 
lattice spacings, the radial and tangential shear speeds will be
different, perhaps by a factor of two
\citep{pp:liquid}. Additionally, the crust may not be in a relaxed
state, causing the background strain to influence the torsional
oscillations. We have not taken this into account in our model.


In the inner crust we expect that at least a portion of the dripped
neutrons are in a superfluid state.  It is, in fact, likely that a
significant portion of the mass does not directly participate in the
oscillations. Of course, the superfluid will nevertheless be set in
motion via the effect of entrainment [for discussion and relevant
references see \cite{ac:sfreview}]. It would be very interesting the
examine the effects of this mechanism on the mode spectrum using the
recently developed formalism of \cite{cs:superelastic} (which
incidentally also allows for a MHD-type magnetic field).

Our models are static, \ie\ non-rotating.  Although this is a very
good approximation for the slowly rotating magnetars it would be
useful to extend our treatment if toroidal oscillations were to be
observed in faster spinning pulsars.

Our model is non-dissipative. In reality a number of processes,
including the emission of gravitational and electromagnetic radiation,
will damp the oscillations. Mutual friction between superfluid and
normal components in the inner crust and viscosity will also be
important. Observations of the QPOs suggest a damping time of a few
tens of seconds in the SGRs. It would be interesting to estimate the
characteristic damping times for each damping mechanism to see if they
are consistent with the observations. In addition, if the emission
mechanism of the observed X-ray signal was known, a lower limit of the
amplitude of the motion of the crust could be derived. This might lead
to an estimate of the maximum strain in the crust and hence an
assessment of the viability of elastic motion as the origin of the
QPOs. Clearly, if the maximum strain exceeds the breaking strain, the
oscillations cannot be of elastic origin.


\section*{Acknowledgements}
We thank Anna Watts for useful discussions and for providing data
before publication. This work was supported by the EU Marie Curie
contract MEIF-CT-2005-009366 (LS).  NA acknowledges support from PPARC
via Senior Research Fellowship no PP/C505791/1.

\appendix
\section[A]{Axial perturbations in the Cowling approximation}
\label{sec:A}
In Newtonian theory, the so-called Cowling approximation 
corresponds to neglecting perturbations in the 
gravitational potential. The accuracy of the approximation depends on the problem
under consideration. For example, while it is not a very good approximation for
the fundamental $f$-mode oscillations of a fluid star it can be excellent 
for the class of gravity $g$-modes which are located in the surface region. 
Basically, the outcome depends on whether the fluid dynamics induces 
significant variations in the gravitational field. Another problem where the 
Cowling approximation is useful is that of inertial modes of rotating stars.    
In this case the dynamics is dominated by the Coriolis force and variations
in the gravitational potential enters at higher orders in the slow-rotation expansion.
 
In this paper we make use of the analogous approximation in the 
framework of general relativity. This problem is obviously somewhat more complex owing 
to the fact that in general relativity the gravitational field
is a dynamical entity described by a tensor field, not just a scalar 
potential. The relativistic Cowling approximation has a long history dating back
to \cite{mvs:cowling} who considered $f$, $p$ and $g$-modes in
neutron stars. Their approach was to simply take the equations 
that had been derived by
Thorne and co-workers, \eg\ \cite{tc:nonradstellar}, and neglect the
metric perturbations. Since the original equations are stated in a specific
gauge it is not clear that this is equivalent to ``throwing away the
gravitational degrees of freedom''. This issue was duly noted by
\cite{finn:cowling} who reconsidered the problem and suggested a different 
form of the approximation. However, even then the gauge problem was
not considered in detail. It was later suggested by \cite{ls:cowling} that
the original formulation gave better results (compared to fully
numerical work) than the modified version for the low order $p$-modes. 
One has to treat this comparison with some caution, however, because Finn's argument 
is mainly relevant for the $g$-modes.  

In this paper we consider torsional oscillations in the crust (as
opposed to the polar perturbation case discussed above). We shall pay special
attention to the gauge problem and define our approximation as 
neglecting the \emph{coupling} between the gravitational and matter degrees of
freedom rather than just setting the
variations of the gravitational field to zero. It would be interesting
to consider this approach also in the polar case  (\eg\ using the
formalism of \cite{gs:gi}) in order to compare with
the work cited above.

\subsection{Axial perturbations in the decoupling limit}

In this appendix we make contact with recent work on relativistic elasticity theory. 
By default this means that we will have to be somewhat technical. 
However, as the complete problem formulation and all astrophysically relevant results
are provided in the main text, the material provided here is 
only intended for readers that are interested in the technical details.

The main results (\ie\ the perturbation equations) that we will describe agree
with previous work in the relevant limits [see
\cite{st:torsional,yl:nonrad,mps:qpo,sks:torsional}], but we extend the results in
the literature to include, in principle, anisotropic and strained
backgrounds. The formalism is also valid, with minor alterations, for
non-static backgrounds.

We assume that the full perturbed spacetime is axisymmetric with the
Killing vector generating the symmetry denoted by $\eta^a$ (we use lowercase Latin letters
to represent spacetime indices). Since the
background around which we will later linearise is assumed to be
spherically symmetric this only implies trivial restrictions of
generality. We write the full spacetime metric as
\beq
  g_{ab} = \perp_{ab} + F\mu_a\mu_b, \qquad \mu_a=F^{-1}\eta_a, \qquad \eta^a\!\!\perp_{ab}=0,
\eeq
where $F=\eta^a\eta_a=r^2\sin^2\theta$ is the norm of the Killing
vector. The tensor field $\perp_{ab}$ is a metric on the manifold of
Killing vector flow lines although it is generically only a metric on a
submanifold of the background spacetime. 
\cite{karlovini:axial} has shown that under these circumstances the axial perturbations with 
arbitrary matter sources are governed by the
gauge invariant set of equations\footnote{The result is, in fact, more
general than this, see \cite{karlovini:axial} for details.}
\begin{align}
  \nabla_b(FQ^{ab}) &= \kappa J^a, \lbeq{Qeq} \\
  \nabla_a J^a &= 0, \lbeq{Jeq}
\end{align}
where $\kappa=8\pi G/c^4$ ($=8\pi$ in geometric units) is the coupling
constant in Einstein's equations,
\beq
 Q_{ab} = 2\nabla_{[a}\delta\mu_{b]},
\eeq
encodes the metric perturbations and
\beq
  J^a = 2\delta(\perp^{ab}\eta^cT_{bc}), 
\eeq
where $T_{bc}$ is the stress-energy tensor, 
describes matter perturbations. In this section the symbol $\delta$ is
used to denote a perturbed quantity in an arbitrary gauge. As discussed
by \cite{karlovini:axial} the axial gauge transformations are
generated by a vector field $\zeta^a$ given by
\beq \lbeq{gaugevec}
  \zeta^a = f_G\eta^a, \quad \eta^a\nabla_a f_G = 0,
\eeq
leading to
\beq\lbeq{gaugetrf}
  \delta \mu_a \rightarrow \delta \mu_a + \nabla_a f_G.
\eeq
If we assume that the metric degrees of freedom are weakly coupled to
matter (which should be a reasonable approximation in the relatively
tenuous crust of a neutron star) we may make the approximation that
$\kappa\rightarrow 0$ in the perturbation equations (we obviously 
 still keep the gravitational constant for the background). We are
then left with decoupled equations for the gravitational waves and
the matter current $J^a$. 

Note that the gauge invariant two-form $Q_{ab}$ is given solely by the
geometric perturbations. Hence, to the extent that the coupling may be
ignored, the gravitational-wave degrees of freedom (\eg\ the $w$-modes
of the star) can be described in a gauge invariant manner. We refer to
equation \refeq{Qeq} with the right hand side set to zero as the
``inverse Cowling'' approximation, cf. \cite{aks:ica}. In order to
clarify the content of this equation, let us write it out in
Regge-Wheeler coordinates. Assuming a harmonic time dependence and
separating out the angular dependence in the standard manner we arrive
at
\beq
  \left(\frac{\der^2}{\der\rstar^2} + \omega^2\right) Z = VZ,
\lbeq{Aperteq}\eeq
where
\beq
  V = 2\frac{1}{r^2}\left(\frac{\der r}{\der \rstar}\right)^2 - \frac{1}{r}\frac{\der^2 r}{\der \rstar^2} 
    +e^{2\nu}\frac{(l-1)(l+2)}{r^2} 
    = \frac{e^{2\nu}}{r^3}\left[l(l+1)r + \frac{\kappa}{2}(\rho-p_t)-6m(r)\right],
\eeq 
and
\beq
  \frac{\der r}{\der \rstar} = e^{\nu-\lambda}.
\eeq
For matter that is unable to support strain (such as perfect fluids)
this is just the standard result for axial $w$-modes
\citep{cf:osc,kokkotas:axialmodes}. If, on the other hand, the
matter is able to support stresses, Eq.~\refeq{Aperteq} generalises
the $w$-mode problem in the inverse Cowling approximation to arbitrary
matter. Note that the shear modulus enters only via the necessary
distinction of pressures, and hence via the background equations. Any
direct appearance is removed by the inverse Cowling
approximation. 

The matter current $J^a$ on the other hand depends on the perturbed
metric and is therefore not as nicely decoupled. Having defined the
``gauge invariant''\footnote{The quotation marks should be taken as a
reminder that, since we are not using the full linearised equations,
the solutions that we obtain do not consistently approximate a
one-parameter solution to the full Einstein equations to the
prescribed order. It is therefore somewhat dubious to discuss 
gauge invariant identifications of points in the perturbed spacetime to
the corresponding points in the background.} inverse Cowling approximation it is
natural to demand that $Q_{ab}=0$ in the Cowling case. Because of
\refeq{gaugetrf} it is clear that this amounts to letting $\delta
\mu_a$ be at most a gradient of a scalar (gauge) function.  The matter
equations will clearly depend on the type of matter. Here we will only
consider conformally deforming perfectly elastic matter
\citep{ks:relasticityI,cq:elastica}, although the generalisation to 
arbitrary matter is straightforward.

\subsection{Elastic matter}

For elastic matter the current takes the form \citep{ks:relastaxial}
\beq
  J^a = 2(\rho+p_t)FS^{ab}\nabla_b (\delta\tilde\phi-f_G),
\eeq
where $p_t$ is the pressure in the directions tangential to the
surfaces of spherical symmetry, $\delta\tilde\phi$ is the perturbed
$\phi$-coordinate on a reference space keeping track of the relaxed
matter configuration [see \eg\ \cite{ks:relasticityI,cq:elastica}]
and $S^{ab}$ is a metric tensor field that describes axial shear-wave
propagation,
\beq
  S^a{}_b = \mrm{diag}(-1, v^2_{r}, v^2_{t}, 0),
\eeq
where $v^2_{r}$ and $v^2_{t}$ are the speed of shear waves in the
radial and tangential direction, polarised in the orthogonal
tangential direction, respectively. In the isotropic limit these
speeds are equal and are given by
\beq
  v^2 = \frac{\check\mu}{\rho + p},
\eeq
where $\check\mu$ is the shear modulus. Under a gauge transformation
generated by the vector field \refeq{gaugevec} we have
\citep{ks:relastaxial}
\beq
  \delta\tilde\phi \rightarrow \delta\tilde\phi + f_G,
\eeq
which means that, in any gauge, we have 
\beq
  J^a = 2(\rho+p_t)FS^{ab}\nabla_b \delta\tilde\phi,
\eeq
whenever $Q_{ab}=0$. Therefore \refeq{Jeq} is also ``gauge invariant''.

An intuitive approach to the relativistic Cowling approximation is to
start out with the conservation equation for the stress-energy tensor
and drop the perturbed metric, obtaining
\beq \lbeq{DdTab}
  \delta(\nabla^a T_{ab})=\nabla^a \delta T_{ab}=0.
\eeq
From the results of \cite{ks:relastaxial} it is easy to see that the
axial part of the perturbed stress-energy tensor, dropping the
perturbed metric, is given by
\beq \lbeq{dTab}
  \delta T_{ab} = 2(\rho+p_t)\eta_{(a}S_{b)c}\nabla^c\delta \tilde\phi. 
\eeq
It is now straightforward to convince oneself that the equation
resulting from \refeq{Jeq} is identical to the projection of
\refeq{DdTab} along $\eta^a$. All other components vanish identically.
Hence, our geometrical way of deriving the Cowling approximation
agrees with the intuitive approach. The advantage we have gained 
is that we now have a clear picture of how gauge issues arise. Most 
importantly, we have 
qualified the approximation as neglecting the coupling between geometrical
degrees of freedom and matter degrees of freedom. This strategy could prove
useful in other contexts, \eg\ in a derivation of  
the Cowling approximation for polar perturbations.

We now proceed to derive explicit equations for our elastic crust. We
start by separating out the angular dependence. Assuming also a
harmonic time dependence we write $\delta\tilde\phi=e^{i\omega
t}C(\theta)F(r)$, where $C(\theta)$ turns out to be given by a
Gegenbauer polynomial, see \cite{karlovini:axial}. The function $F$
has the same meaning as $Y$ in \cite{st:torsional}, $T$ in
\cite{yl:nonrad}, $W/r$ in \cite{mvh:nonrad} and $F$ of
\cite{mps:qpo} if the relevant limits are taken.
In Schwarzschild coordinates, denoting derivatives with respect to the
Schwarzschild radius $r$ by a prime, we find
\begin{align}\lbeq{perteqA}
  F'' + A'F' + BF = 0,
\end{align}
where
\begin{align}
  e^A &= r^4e^{\nu-\lambda}(\rho+p_t)v_{r}^2 ,\\
  B &= \frac{e^{2\lambda}}{v_{r}^2}\left[e^{-2\nu}\omega^2-\frac{v_{t}^2(l-1)(l+2)}{r^2}\right].
\end{align}

These equations need to be complemented by boundary conditions at the top 
and bottom of the crust region. These conditions are naturally taken to be that the
traction $t_a = r^b\delta T_{ab}$ vanishes at the boundaries. We then have 
\beq
  t_a = (\rho+p_t)\eta_av_{r}^2r_b\nabla^b \delta\tilde\phi =0.
\eeq 
That the traction should vanish follows from the fact that the shear modulus,
and hence the shear wave speed, is zero in the fluid or vacuum
surrounding the solid. In terms of $F$ we get
\beq
  t = r^2F' = 0,
\eeq
at the boundaries. We now define a rescaled traction according to
$T=e^Ar^{-2}t=e^A F'$ and use this to reduce \refeq{perteqA} to a first
order system,
\begin{align}\lbeq{peqsysA}
  T' &= -e^ABF, \\
  F' &= e^{-A}T.
\end{align}
This system is convenient for numerical integration since there are no
derivatives of the equation of state parameters such as the shear
modulus and the density (which are only known in tabular form). When the
background is isotropic, {\em i.e.,} when the two shear-wave speeds are identical
$v_r =v_t =v$, the system simplifies considerably and one
obtains
\begin{align}
  e^A &= r^4e^{\nu-\lambda}\check\mu, \\
  B &= e^{2\lambda}\left[\frac{e^{-2\nu}\omega^2}{v^2} - \frac{(l-1)(l+2)}{r^2}\right].
\end{align}
The boundary conditions are now simply that $T=0$ at the top and bottom of the crust.

One can check that this system is equivalent to that obtained by
\cite{yl:nonrad}, which has the correct Newtonian limit
\citep{mvh:nonrad}. The equations of \cite{st:torsional} and \cite{mps:qpo} also reduce to these
equations 
when the metric perturbations or magnetic fields are set to zero, respectively.

\section{The crust thickness}
\label{sec:B}

\begin{figure}
\centerline{\includegraphics[width=0.45\columnwidth,angle=0,clip]{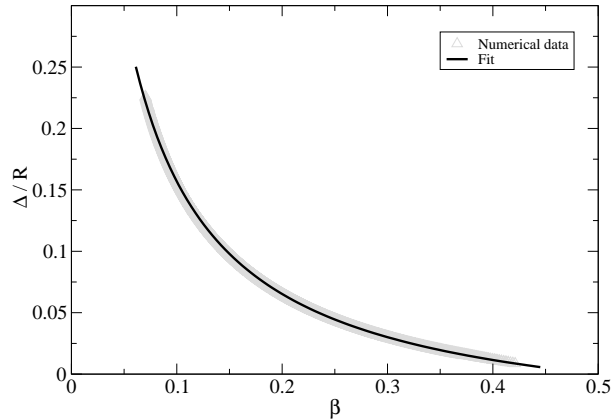}}
\caption{Fit of $\Delta/R$ with $\alpha=0.02326$ to numerical results for a large sample of stellar models.
Note especially that $\Delta/R$ is (very nearly) 
a function of $\beta$ only, and that the simple one-parameter fit suggested by the polytropic
 approximation is quite accurate.}
\label{fig:Delta}
\end{figure}

In order to express the estimated crust mode frequencies in terms of
just $M$ and $R$ we need to find the crust thickness as a function of
these variables. To this end we approximate both the mass function and
the metric function $\lambda$ as being constant and given by
$e^{-2\lambda}=1-2\beta$, where $\beta=M/R$. We also use the fact that
the pressure is negligible compared to $\rho$ and $\beta$ in the crust
and take the equation of state to be a simple polytrope,
\beq
  \rho = k p^{1/\Gamma},
\eeq
with $k$ and $\Gamma$ constant.  With these approximations the
equation of hydrostatic equilibrium becomes
\beq
  p' \approx -\rho \frac{M}{r^2} e^{2\lambda},
\eeq
which may be solved to yield
\beq \lbeq{presseq}
  \chi p^{1/\chi} \approx ke^{2\lambda}\frac{M}{r} + \mathcal{C},
\eeq
where $\mathcal{C}$ is an integration constant and
$\chi=\Gamma/(\Gamma-1)$. Denoting the pressure at the crust-core
interface (at $r=R_c$)  by $p_c$ (and defining a
corresponding transition density $\rho_c$) we solve for $\mathcal{C}$ to find
\beq
  \mathcal{C}=\alpha-\frac{M}{R_c}e^{2\lambda},
\eeq
where $\alpha=\chi p_c/\rho_c$. Putting this into \refeq{presseq} and
setting $p(R)=0$ at the surface we obtain
\beq
  e^{2\lambda}M\left(\frac1{R}-\frac1{R_c}\right)+\alpha \approx 0,
\eeq
solving for $\Delta/R=(R-R_c)/R$ we finally obtain
\beq \lbeq{drrel}
  \frac{\Delta}{R} \approx \left(\frac{\beta}{\alpha}e^{2\lambda}+1\right)^{-1}. 
\eeq
We see that this is a function of the compactness $\beta$ only, something we have
confirmed in the numerical study, see Figure~\ref{fig:Delta}. 
The value of $\alpha$ obviously depends on the equation of state. For
our tabulated equation of state it can be written
\beq
  \alpha \approx 0.0047\chi_{\mrm{eff}},
\eeq
where $\chi_{\mrm{eff}}$ is an effective value. 
For a $\Gamma=4/3$ polytrope we obtain $\alpha\approx
0.019$. We also fitted the expression \refeq{drrel} using $\alpha$ as
a free parameter to the numerically obtained data and found
\beq
  \alpha \approx 0.02326,
\eeq
agreeing well with the estimated value. This value corresponds to
$\chi\approx 5$ or $\Gamma\approx \frac54$. 
In our detailed numerical fits described in the main text we use 
\refeq{drrel}. However, if one is satisfied with an accuracy at the 10\% level then 
the simple approximation
\beq 
  \frac{\Delta}{R} \approx \frac{\alpha}{\beta}e^{-2\lambda},
\eeq
is relevant for all neutron stars (since $M/R\gg2\times10^{-2}$). 



\end{document}